\documentstyle[epsfig,preprint,aps]{revtex}
\begin{document}
\draft
\def\bra#1{{\langle #1{\left| \right.}}}
\def\ket#1{{{\left.\right|} #1\rangle}}
\def\bfgreek#1{ \mbox{\boldmath$#1$}}
\title{The neutron charge form factor in helium-3}

\author{D.H. Lu, K. Tsushima, A.W. Thomas and A.G. Williams}
\address{Department of Physics and Mathematical Physics\break
 and
        Special Research Centre for the Subatomic Structure of Matter,\break
        University of Adelaide, Australia 5005}
\author{K. Saito}
\address{Physics Division, Tohoku College of Pharmacy,\break
Sendai 981-8558, Japan}
\maketitle

\vspace{-9.3cm}
\hfill ADP-98-07/T286
\vspace{9.3cm}
\begin{abstract}
In order to measure the charge form factor of the neutron, $G^n_E(Q^2)$, 
one needs to use a neutron bound in a nuclear target. We calculate the change
in the form factor for a neutron bound in $^3He$, with respect to the free
case, using several versions of the quark meson coupling model.  
It is found that the form factor may be suppressed by as much as 12\% 
at $Q^2 = 0.5\mbox{ GeV}^2$ with respect to that of the free neutron. 
\end{abstract}

\pacs{PACS numbers: 12.39, 21.65, 13.40.Gp}


There are currently a number of very sophisticated experiments underway
at Bonn, Mainz, MIT Bates and TJNAF, in which it is proposed to
measure the elusive, electric form factor of the neutron, $G^n_E(Q^2)$
\cite{drechsel97,he3,deuteron}. This quantity is of great interest in
QCD because of the subtle interplay of quark and meson degrees of
freedom\cite{IK,CBM}. In the absence of a free neutron target, the
neutron form factors have to be extracted from electron-nucleus
scattering. However, because it has zero charge the neutron charge form
factor vanishes at $Q^2 = 0$ and is substantially smaller than the
proton charge form factor for $Q^2 < 1 \mbox{ GeV}^2$.
As a consequence, the extracted neutron charge form factor 
has typically been associated with large systematic errors -- 
except for its charge radius, which can be measured
accurately by the scattering of ultra-cold neutrons by 
electrons\cite{rms}. By using a polarized beam and target the new  
experiments aim to produce results which should be much more direct and 
quite insensitive to two-body exchange currents\cite{Arenhovel87}.

Our concern is that the binding energy of $^3He$ (and the central
density) is significantly higher than for the deuteron and therefore the
possible medium modification of the internal structure of the bound 
``nucleon'' could be significant. We therefore investigate the
modification of the bound neutron charge form factor in $^3He$
within the quark-meson coupling (QMC)
model\cite{Guichon88,Tony94,Guichon96}. This effect has been estimated before.
For example, it has been shown that the neutron charge form 
factor is altered dramatically at small momentum transfer in the chiral 
$\pi\rho\omega$ Lagrangian\cite{Meissner89}, 
the linear sigma model\cite{Goeke89}
and the chiral bag model\cite{Cheon92}.
In Ref.\cite{inmedium97}, we showed that the in-medium nucleon 
electromagnetic form factors in the standard QMC model are somewhat reduced, 
and they are consistent with constraints deduced from 
$y$-scaling data\cite{sick}. 

Recently, extensions of QMC have been proposed in which, in order to 
simulate the eventual quark-gluon deconfinement at high density, the bag
constant is allowed to decrease as the bag density (or mean scalar field)
increases\cite{JJ96,Blunden96,bagconst}. Here we examine the variation of 
$G_E^n(Q^2)$ for a neutron bound in $^3He$ for several of these variations 
that are consistent with other constraints\cite{sick}, as well as for 
the standard QMC model.

For convenience we briefly review the standard version of QMC in which
the bag constant, $B$, is fixed.
The additional complications when $B$ varies with density are
explained in Ref.~\cite{bagconst}.
At the present stage of development of the QMC model, 
a nuclear system is viewed as a collection of non-overlapping 
MIT bags\cite{MIT}. The medium effects arise through the self-consistent
coupling of phenomenological scalar $(\sigma)$  and vector $(\omega,\rho)$ 
meson fields to the confined valence quarks --- rather than
to the nucleons, as in Quantum Hadrodynamics (QHD)\cite{QHD}. 
As a result, the internal structure of the bound ``nucleon'' 
is modified by the medium with respect to the free case.

In mean field approximation for the meson fields, the lowest state of the 
quark inside a bag with a radius R is given by\cite{MIT} 
\begin{equation}
q_m(t,{\bf r}) =  {N_0\over\sqrt{4\pi}}  e^{-i\epsilon_q t/R} \left (
\begin{array}{c} j_0(x r/R) \\ 
i\beta_q \bfgreek {\sigma} \cdot\hat{r} j_1(x r/R) \end{array} \right )
\, \theta(R-r)\, \chi_m , \label{cavity}
\end{equation}
where 
\begin{eqnarray}
\epsilon_q {u\choose d} 
&=& \Omega_q + g_\omega^q\overline{\omega} R 
\pm{1\over 2}g_\rho^q\overline{b}, \hspace{1cm}
\beta_q = \sqrt{\Omega_q-m_q^*R \over \Omega_q + m_q^*R}, \\
N_0^{-2} &=& 2R^3 j_0^2(x)[\Omega_q(\Omega_q-1)+ m_q^*R/2]/x^2,
\end{eqnarray}
with $\Omega_q \equiv \sqrt{x^2 + (m_q^*R)^2}$, 
$m_q^* \equiv m_q - g_\sigma^q\overline{\sigma}$, 
and $\chi_m$ the quark Pauli spinor. The parameters in this expression 
($x$, $m_q^*$ and $R$)
are determined by the boundary condition at the bag surface,
$j_0(x) = \beta_q j_1(x)$,
together with the mean values of the scalar ($\overline{\sigma}$)
and the time components of the vector ($\overline{\omega}$ and $\overline{b}$)
fields ($\overline{b}$ stands for the neutral $\rho$ meson mean field),
which are governed by the equations of motion,
\begin{eqnarray}
\overline{\omega} &=& {g_\omega\rho\over m_\omega^2}, \label{omega}\\
\overline{\sigma} &=& 
{g_\sigma\over m_\sigma^2} C(\overline{\sigma}){2\over (2\pi)^3}
\sum_{i=p,n}\int^{k_F^{(i)}} d^3 k {m_N^*(\overline{\sigma})\over 
\sqrt{m_N^{*2}(\overline{\sigma}) + k^2}}, \label{scc}\\
\overline{b} &=& {g_\rho\over 2m_\rho^2} (\rho_p - \rho_n).
\end{eqnarray}
Here $\rho=\rho_p + \rho_n$  is the baryon  density, 
$k_F^{p(n)}$ is the proton(neutron) Fermi momentum, 
the bound nucleon effective mass is
$m_N^*(\overline{\sigma}) = {3\Omega_q / R}
- { z_0/ R} + (4\pi/3) R^3 B_0$,
and the function $C(\overline{\sigma})$ is $S(\overline{\sigma})/S(0)$, with
$S(\overline{\sigma}) = 
[\Omega_q/2+m_q^*R(\Omega_q-1)]/[\Omega_q(\Omega_q-1)+m_q^*R/2]$. 
The meson-nucleon coupling constants are related to the meson-quark coupling
constants by $g_\sigma= 3g_\sigma^qS(0)$, $g_\omega= 3g_\omega^q$, and 
$g_\rho = g_\rho^q$.

In practice, the quantities $z_0$ and $B_0$
are first determined by requiring the free nucleon  mass to be 
$m_N^*(\overline{\sigma}=0) = 939 $ MeV  and by the stability condition, 
${\partial m_N^*(\overline{\sigma}) / \partial R} = 0$,
for a given bag radius of the free nucleon, $R_0$ 
(treated as an input parameter).
Then the coupling constants, $g_\sigma$, $g_\omega$ and $g_\rho$, are
chosen to fit the saturation properties of nuclear matter 
(saturation energy and bulk symmetry energy) at normal nuclear density.  
Note that the $\rho$ meson is now explicitly included in addition to the 
original $\sigma$ and $\omega$ meson fields in order to describe  
asymmetric nuclear matter.
These  coupled, nonlinear equations  can be solved self-consistently for
an arbitrary baryon density, where
the solution contains a medium-modified quark wavefunction.
 
The neutron charge form factor extracted from the electron--nucleus scattering
experiment should be interpreted as an average value of the neutron 
contribution over the finite nucleus. 
In QMC, the neutron substructure is modified by its surrounding 
nuclear medium. In a local density approximation, the neutron charge
form factor in a finite nucleus can be written as
\begin{equation}
G_{E}^n(Q^2)= \int G_{E(\rm{n.m.})}^n (Q^2,\rho(\vec{r})) 
\rho_n(\vec{r}) \,d\vec{r},
\end{equation}
where $G_{E(\rm{n.m.})}^n (Q^2,\rho(\vec{r}))$
denotes the density-dependent charge form factor of a neutron immersed in a
uniform density of protons ($\rho_p$) and neutrons ($\rho_n$).

We neglect other residual off-shell effects of the 
bound nucleon\cite{Deforest83}, and so the neutron charge form factor 
can be conveniently evaluated in the Breit frame,
\begin{eqnarray}
G_E^n(Q^2)&=& \bra{n({\vec{q}/ 2})}j^0(0)\ket{n(-{\vec{q}/ 2})}, \\ \label{GE} 
j^\mu(x) &=& \sum_f e_f \overline{q}_f(x) \gamma^\mu q_f(x) 
	 -i e [ \pi^\dagger(x) \partial^\mu \pi(x)
               -\pi(x) \partial^\mu \pi^\dagger(x)], \\ \label{current}
\ket n &=& \sqrt{Z_2^n} [ 1 + (m_n - H_0 - \Lambda H_I \Lambda )^{-1} H_I ] 
\ket {n_0},  \label{state}
\end{eqnarray}
where  $Q^2 = -q_\mu^2 = \vec{q}^{\,2}$, $q_f(x)$ is the quark field operator 
of the flavor $f$, $e_f$ is its charge operator, $\pi(x)$ either destroys 
a negatively charged pion or creates a positively charged one,
$\Lambda$ is a projection operator which projects out all the components
of $\ket n $ with at least one pion, $H_0$ is the Hamiltonian for the bare
baryon and free pion, and $H_I$ is the interaction Hamiltonian
which describes the process of emission and absorption of pions.
The probability of finding a bare neutron bag, $\ket{n_0}$, 
in the physical neutron, $Z^n_2$, is determined by the probability 
conservation condition\cite{CBM}.

To calculate the neutron form factor, we use the Peierls-Thouless 
projection method\cite{PT62}
combined with a Lorentz contraction for the nucleon internal 
wave function in the Breit frame\cite{Lu97,LP70}. 
These techniques for implementing the center-of-mass and
recoil corrections have been quite successful in the case of the 
electromagnetic form factors for free nucleons\cite{Lu97}.
Detailed calculations give
\begin{eqnarray}
G_E^{n(\rm{quark})}(Q^2) &=&  \left( {2\over 3} P_{NN} - 
{1\over 3} P_{\Delta\Delta} \right) G_E^{(\rm{core})}(Q^2), \\
G_{E}^{n(\rm{pion})}(Q^2) &=& G_E^{(\pi)}(Q^2; N) +  
                         G_E^{(\pi)}(Q^2; \Delta), 
\end{eqnarray}
where 
\begin{eqnarray}
P_{BC} &=&  {f^{NB*}f^{NC*}\over 12\pi^2 m_\pi^2}
\int_0^\infty {dk\, k^4 u^2(k R)\over 
(\omega_{BN}+ \omega_k) (\omega_{CN}+\omega_k)\omega_k}, \\
G_E^{(\pi)}(Q^2;N) &=& -{1\over 36\pi^3}\!
\left ({f^{NN*}\over m_\pi}\right )^2 
\!\!\int\! d^3k\, {u(kR)\, u(k' R)\, \vec{k}\cdot\vec{k'}\over
\omega_k \omega_{k'} (\omega_k + \omega_{k'}) }, \label{EN}\\
G_E^{(\pi)}(Q^2;\Delta) &=& 
{1\over 72\pi^3}\!\left ( {f^{N\Delta *}\over m_\pi}\right )^2
\!\!\int\! d^3k\, {u(kR)\, u(k' R)\, \vec{k}\cdot \vec{k'}\over
(\omega_{\Delta N} + \omega_k) (\omega_{\Delta N} + \omega_{k'}) 
(\omega_k + \omega_{k'})}. \label{ED}
\end{eqnarray}
Here $u(kR)=3j_1(kR)/kR$, $\omega_k=\sqrt{{m_\pi^*}^2 + \vec{k}^2}$,
$\omega_{BN}\simeq m_B^* - m_N^*$, $\vec{k}' = \vec{k} + \vec{q}$,  and
$f^{NB*}$ is the renormalized $\pi NB$ coupling constant in medium.
The charge form factor for the quark core of the proton is
\begin{eqnarray}
G_E^{\rm{core}}(Q^2) &=& \int\!d^3r j_0(Qr)f_q(r)K(r)/D_{\rm PT},\label{PTE}\\
D_{\rm PT} &=& \int\! d^3r f_q(r) K(r),
\end{eqnarray}
where 
$f_q(r) \equiv j_0^2(x r/R) + \beta_q^2 j_1^2(x r/R)$ and 
$K(r) \equiv \int\! d^3z \, f_q(\vec{z}) f_q(-\vec{z} - \vec{r})$
is the recoil function to account for the correlation of the 
two spectator quarks.
The effect of the Lorentz contraction for the quark core is taken into account 
by a simple rescaling formula\cite{Lu97}, 
$G_E(Q^2) = ({m_N^*/ E^*})^2 G^{\rm sph}_E(Q^2 {m_N^*}^2/{E^*}^2)$, 
where $E^*=\sqrt{{m_N^*}^2 + Q^2/4}$ and 
the superscript ``sph'' refers to the form factor calculated with the 
spherical bag wave function.

The pion cloud plays a vital role  in this calculation.
For the simplest case where the valence quarks in the bare neutron bag,
$\ket{n_0}$, occupy the $1s$ state in the bag, the final small but 
non-vanishing $G_E^n(Q^2)$ comes primarily from
the photon interaction with either the proton
or the pion in the virtual process, $n\rightarrow p\pi^-$.

In principle, the existence of the $\pi$ and $\Delta$ inside the nuclear 
medium will also lead to some modifications of their properties.
As the $\Delta$ is treated on the same footing as the nucleon,
its mass should vary in a similar manner as the nucleon. Thus,
as a first approximation, we assume that the in-medium and free space 
$N-\Delta$ mass splittings are
approximately equal, i.e., we take 
$m_\Delta^* - m_N^* = m_\Delta - m_N$.
As the pion is  a nearly perfect Goldstone boson, 
%
%
we also take $m_\pi^* = m_\pi$ in this work. 
The $\pi NN$ coupling constant in free space is taken to be the empirical 
value, i.e. $f^{NN} \simeq 3.03$, which corresponds to the usual $\pi NN$ 
coupling constant, $f^2_{\pi NN}\simeq 0.081$. 
In the medium, the $\pi NN$ coupling constant
might be expected to decrease  slightly because of the enhancement of the 
lower component of the quark wave function. It can be shown that 
this reduction is about one-third of that of the $\sigma NN$ coupling 
constant, $g_\sigma(\overline{\sigma})$, which behaves as
$g_\sigma(\overline{\sigma})/g_\sigma\simeq 1-a(g_\sigma\overline{\sigma})/2$. 
Hence, $f^{NN*}/f^{NN}\simeq 1-a(g_\sigma\overline{\sigma})/6$,
where $a \simeq (8.8,11)\times 10^{-4}\mbox{ MeV}^{-1}$ 
for $R_0=(0.8,1.0)$ fm\cite{Tony94}
and $g_\sigma\overline{\sigma}\simeq 
s_1 (\rho/\rho_0) + s_2(\rho/\rho_0)^2 + s_3(\rho/\rho_0)^3$ 
for QMC-II --- c.f. Ref\cite{Saito97}, 
where the coefficients $s_i$ may be found. 

\begin{figure}
\vspace{2.5cm}
\centering{\
\epsfig{file=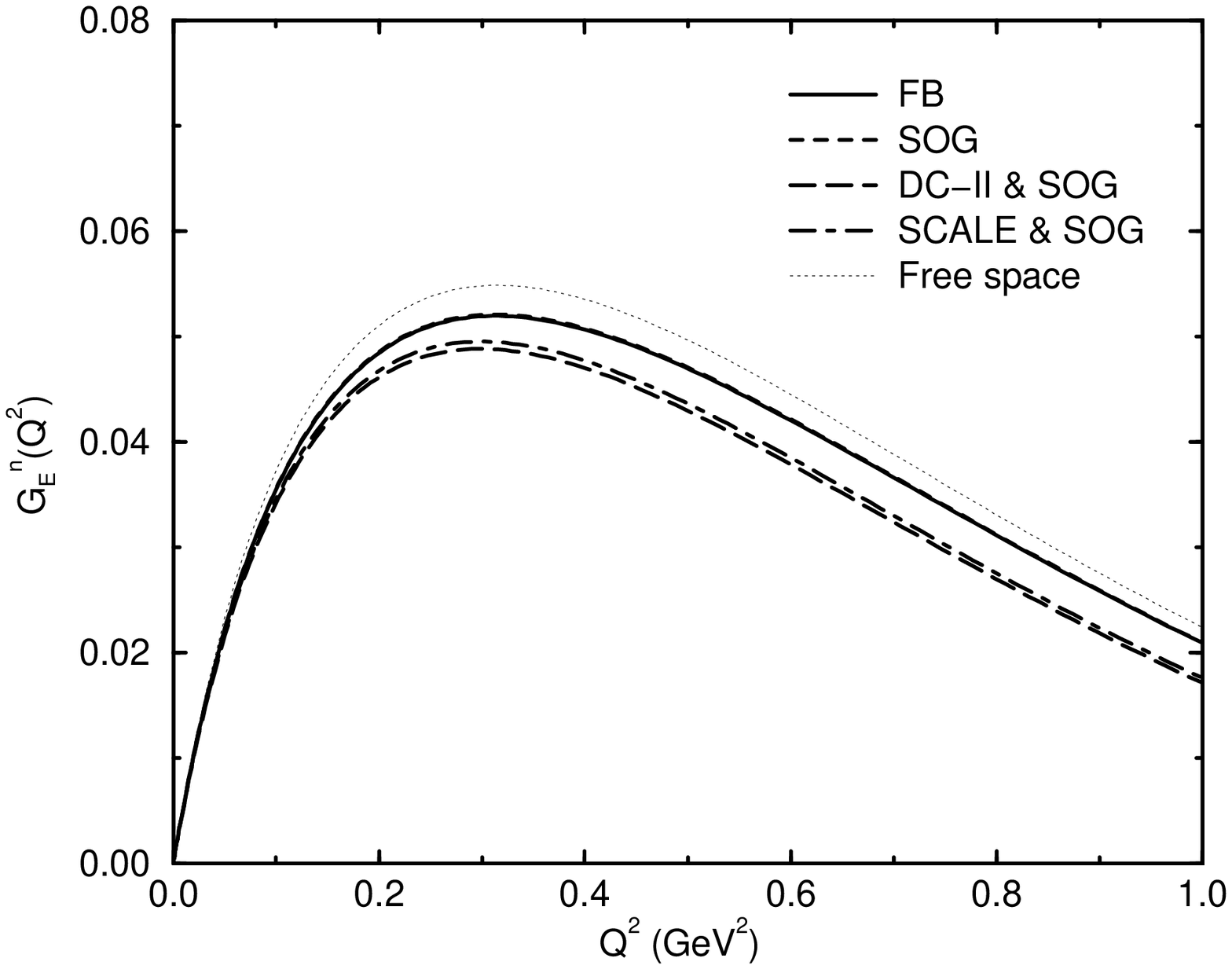,width=12cm}
\vspace{1cm}
\caption{The neutron charge form factor in $^3He$ using the standard QMC model
and two different density parametrizations (FB and SOG stand for a 
Fourier-Bessel fit and a Sum-of-Gaussians fit). The effect of a possible 
reduction of the bag constant is also shown. The $\pi NN$ coupling constant 
is maintained at its free space value. Here DC-II and SCALE refer to two ways 
of implementing a reduction of the bag constant (see text). 
The corresponding free space neutron charge form factor is also shown
for ease of comparison. 
(The free bag radius was taken to be $R_0=1.0$ fm.)}
\label{fig1.ps}}
\end{figure}

There is no direct experimental data for the neutron density distribution 
in $^3He$. We identify the neutron density 
distribution in  $^3He$ with the proton density in $^3H$, which is 
certainly reasonable provided that the charge symmetry breaking is small. 
The baryon density distribution in $^3He$ is 
calculated from the charge densities of $^3He$ and $^3H$. 
Fig.~\ref{fig1.ps} shows the neutron charge form factor with $R_0=1.0$ fm,
using two different experimental fits to the charge distribution 
(FB refers to a Fourier-Bessel fit\cite{FB} and 
SOG to a fit using a Sum-of-Gaussians\cite{SOG}). 
The $\pi NN$ coupling constant is taken to have its free space value in these
calculations. 
The standard QMC gives roughly a 5\% reduction with respect to the free case
at $Q^2=0.5\mbox{ GeV}^2$. The long-dashed and the dot-dashed curves
in Fig.~\ref{fig1.ps} are for variations of the original QMC involving  
possible reduction of the bag constant\cite{JJ96,Blunden96,bagconst}.
The SOG fit is used for these two curves.
DC-II refers to a QMC variant where the bag constant is directly related to 
$\overline{\sigma}$ through
$B/B_0= e^{-4g_\sigma^B\overline{\sigma}/m_N}$ and SCALE to a model
where $B/B_0=(m_N^*/m_N)^{\kappa}$.  
According to $y$-scaling constraints\cite{sick},
the maximum values allowed for the parameters $g_\sigma^B$ and $\kappa$ 
would be 1.2\cite{bagconst}. 
It is clear that the possible reduction of the neutron charge form factor 
is significant in QMC once the bag constant is allowed to decrease
--- i.e., the reduction is nearly 12\% at $Q^2=0.5 \mbox{ GeV}^2$, for both 
parametrizations of the bag constant.

\begin{figure}
\centering{\
\epsfig{file=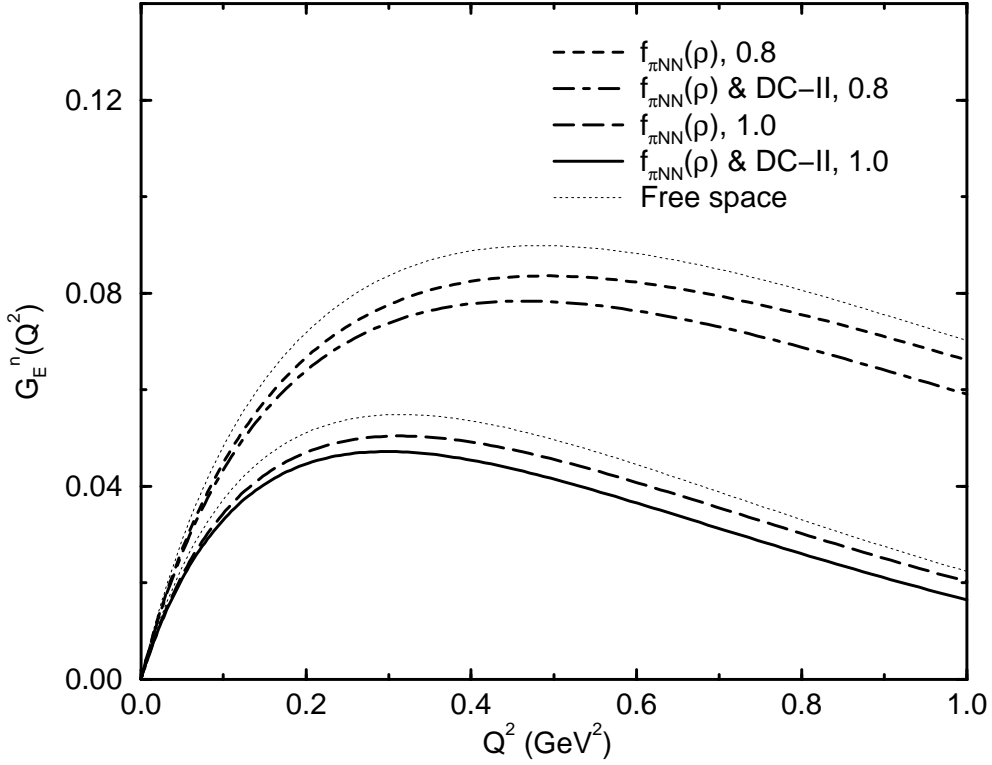,height=12cm}
\caption{The neutron charge form factor in $^3He$ with a possible reduction
of the $\pi NN$ coupling constant in medium. 
The numbers in the legends indicate the free bag radius (in fm) used in each 
calculation. The dashed and long-dashed curves refer to the standard QMC 
with density-dependent $\pi NN$ coupling constant; and the dot-dashed and 
solid lines refer to a QMC with the density-dependent $\pi NN$ coupling 
constant, together with the maximum allowed reduction of the bag constant 
in the direct coupling model. The dotted curves are for the corresponding 
form factors in free space for both bag radii.}
\label{fig2.ps}}
\end{figure}

Fig.~\ref{fig2.ps} presents the dependence of the neutron charge form factor
on the $\pi NN$ coupling constant with two different bag radii. 
Note that the absolute value of the neutron charge form factor is clearly 
quite sensitive to the bag radius, however, the really important issue here
is the relative change with respect to the free case. 
For a typical reduction of the $\pi NN$ coupling constant in medium 
(3\% at $\rho_0$ in QMC) with a fixed
bag constant, the neutron charge form factor decreases
by less than 2\% at $Q^2=0.5\mbox{ GeV}^2$. 
If we allow variation in the $\pi NN$ coupling constant and allow the 
bag constant to decrease at the same time, the combined possible reduction for 
the neutron charge form factor is also about 
12\% for $R_0=1.0$ fm (solid line) and 
8\% for $R_0=0.8$ fm (dot-dashed line) at $Q^2=0.5\mbox{ GeV}^2$.

At present the only experimental information on the modification 
of $G_E^n(Q^2)$ comes from the comparison of measurements of the asymmetry
in the scattering of polarized electrons from polarized  $^3He$\cite{he3}
and deuteron\cite{deuteron} targets.
At $Q^2 \simeq 0.35\mbox{ GeV}^2$ the deuteron measurement is significantly
larger than the $^3He$ measurement --- the ratio being almost a factor of two.
While the experimental errors do not allow firm conclusions to be drawn yet,
the situation will improve dramatically in the next few years.

In conclusion, we have calculated the neutron charge 
form factor in $^3He$ using QMC and a number of its variants. 
The in-medium quenching of the neutron charge form factor is insensitive 
to the details of the $^3He$ matter density chosen. The neutron charge 
form factor is the result of a cancellation between the contributions 
of a $\pi^-$ cloud and a proton-like core. Because the long-range pion
cloud is not greatly altered in medium the charge radius decreases by only 
a few percent (3.3\% and 3.8\% for $R_0$ = 1.0 and 0.8 fm, respectively). 
However, the expansion of the core (especially when the
bag pressure is allowed to decrease) does lead to a significant decrease
of $G_E^n(Q^2)$ above $0.2 \mbox{ GeV}^2$. 
The maximal allowed reduction of the neutron charge charge 
form factor at $Q^2=0.5\mbox{ GeV}^2$ 
may reach 12\% for a reasonable bag radius, such as $R_0=1.0$ fm.
The present results are comparable to the previous results in 
the generalized Skyrme model\cite{Meissner89},
the chiral sigma model\cite{Goeke89},
and the chiral bag model\cite{Cheon92}.
It is worthwhile to point out that while the picture of
the nucleon in these models is  quite different, the consistency
of the predictions implies a certain degree of model-independence of 
the results. We would, however,  like to emphasize that our model describes 
the nucleon and nuclear system consistently. 
It is able to reproduce the nuclear charge distributions 
for the closed shell nuclei, as well as
the saturation energy, density and compressibility of nuclear matter.
The medium modification implied by the model is  
consistent with the current experimental limits, and it will be 
very interesting if it is confirmed in future, more precise measurements.

This work was supported by the Australian Research Council.

\newpage
\vspace*{1cm}
\end{document}